\documentstyle[sprocl,epsfig]{article}

\newcommand{\beq}{\begin{equation}}
\newcommand{\eeq}{\end{equation}}
\newcommand{\bea}{\begin{eqnarray}}
\newcommand{\eea}{\end{eqnarray}}

\def\xv{{\vec x}}

\def\qv{{\vec q}}
\def\rv{{\vec r}}
\def\N{{\scriptscriptstyle N}}

\begin{document}
 
\title{THE NEUTRON DISTRIBUTION IN NUCLEI AS MEASURED WITH PARITY 
VIOLATING ELECTRON SCATTERING}

\author{W.M.ALBERICO, A.MOLINARI\thanks{
Travail present\'e au ``Rencontre sur la parit\'e'', 
Ecole Normale Superieure,
Paris, Novembre 1997}}

\address{ Dipartimento di Fisica Teorica, 
via P.Giuria 1 -- 10125 Torino, Italy\\
and\,\,
 I.N.F.N., Sezione di Torino}

\maketitle\abstracts{A short  review of the  present knowledge 
of the nucleons distribution in nuclei is given. A proposal is made about
a possible measurements of the neutron distribution through polarized 
electron scattering off nuclei.
}

\section{The nuclear mean field}

The cross sections involving neutrons and protons 
impinging on nuclei display a marked energy variation 
generally interpreted as an interference 
between the incident and the transmitted waves.
This occurrence  in turn implies a 
mean free path for collisions between the nuclear 
constituents large compared not only to the internucleonic distance, 
but even, sometimes, to the dimensions of the nucleus itself.

This finding is strongly suggestive of a mean field approach to the
structure of the nucleus, especially as far as its ground state is 
concerned.

The nuclear mean field is implemented at the empirical level with 
the shell model. At the theoretical level a natural treatment 
resorts to the Hartree--Fock (HF) self--consistent theory, in a frame
viewing nuclei  as self--bound composite systems of nucleons 
interacting via a static two--body potential and 
governed by non--relativistic quantum mechanics. 

Care is however 
required in handling the HF theory for the atomic nuclei. Indeed any 
realistic nucleon--nucleon (NN) force $V_{NN}$ embodies so much 
repulsion that the expectation value of the nuclear Hamiltonian in the
HF ground state is far from being enough attractive. In other words 
the HF wave function does not prevent two nucleons to come 
close to each other, where they experience a violent repulsion.

The Brueckner theory provides a remedy to this flaw. Indeed it 
yields an effective interaction $G$ between two nucleons in a 
nucleus such that the interaction of two ``uncorrelated'' nucleons 
through $G$ equals the interaction of two ``correlated'' 
nucleons through $V_{NN}$. Formally this result is achieved in the 
framework of perturbation theory by summing up (with the Bethe--Goldstone
equation) the infinite set of ladder diagrams representing the 
scattering of two nucleons in the medium out of the Fermi sea, the 
other nucleons remaining passive.

Is the Brueckner--Hartree--Fock (BHF) accounting satisfactorily 
for the nuclear ground state properties? While the answer to this question 
is negative (the BHF mean field yielding a too large nuclear 
central density), yet an important lesson has been learned from  BHF.
It amounts to recognize that the  perturbative corrections to the BHF 
mean field, necessary to reconcile theory and experiment, 
should not be computed 
by expanding in $G$, but rather by grouping together diagrams involving 
three, four, etc. nucleons repeatedly interacting among themselves 
through $G$ ({\it hole line} expansion).

In spite of the fact that  a proof of the 
convergence of the hole line expansion has never been provided, 
three-- and four--hole line contributions have been actually 
computed in nuclear matter; however to achieve the same goal in finite nuclei
has proved to be an almost impossible task. 

Accordingly short cuts have been seeked  
 to incorporate the effects going beyong BHF still in a 
self--consistent mean field framework for the nuclear ground 
state.\footnote{We do not mention, for brevity, relativistic mean field 
approaches which also look promising in reproducing 
the nuclear ground state properties.}
In this connection a successfull approach allows for a density
dependence of the interaction $G$, in addition to the one naturally
induced by the Pauli operator in the Bethe--Goldstone equation.
Indeed this procedure conveniently simulates 
 the energy dependence and non--locality of $G$. Moreover 
it explains quite successfully the 
experimentally well established fact that in a nucleus the single 
particle orbits below the Fermi level are not 100\% occupied as the
HF (but not the BHF) approach would imply. Clearly the prize to be 
payed for the simplicity (and the success) is the introduction of
a certain amount of  phenomenology empirically fixing the local 
density dependence of $G$. 

In the above outlined scheme, Negele\cite{Negele} has been able
to impressively reproduce over a wide range of momentum transfer
 the data of elastic electron scattering on several nuclei (like 
$^{16}$O, $^{40}$Ca and $^{208}$Pb). This findings strongly support 
the view that most of the physics of the nuclear  ground state lends 
indeed itself to be embodied in a ``mean field''. 

Should this be the case, then we would actually know and understand 
the nucleons' distribution in nuclei. However, before drawing this 
conclusion, one should remind  that
electrons only probe the {\it proton} distribution.
It would be therefore desirable to test the mean field scheme on other
observables, the most natural one being the neutron distribution.

Although experimental information on the latter has been gathered in 
the past, e.g. with pion scattering on nuclei, still our knowledge of it 
remains rather poor. Even the recurring question of whether or not the nuclear
surface is neutron rich cannot be presently answered with certainty. 

Furthermore  the present workshop stresses the urgence of reaching an 
accurate knowledge of the neutron distribution  in order
to achieve a precise interpretation of the atomic parity--violating (PV)
experiments. Thus in the following we propose a method to measure the
neutron distribution in the nuclear ground state, which is based on 
PV polarized electron--nucleus scattering.

\section{The formalism of PV electron scattering}

The helicity asymmetry, as measured in the scattering of rigth-- and 
left-- handed electrons off nuclei, is defined as follows
\bea
{\cal A} &&= \frac{d^2\sigma^+ - d^2\sigma^-}
{d^2\sigma^+ + d^2\sigma^-}
\nonumber\\
&&={\cal A}_0\frac{v_L R^L_{AV}(q,\omega) + v_T R^T_{AV}(q,\omega) 
+ v_{T'} R{T'}_{VA}(q,\omega)}
{v_L R^L(q,\omega) + v_T R^T(q,\omega)}
\label{asym}
\eea
where
\beq
v_L = \left(\frac{Q^2}{\qv^2}\right)^2,
\label{vL}
\eeq
\beq
v_T = \frac{1}{2}\left\vert \frac{Q^2}{\qv^2}\right\vert +
\tan^2\frac{\theta}{2}
\label{vT}
\eeq
and
\beq
v_{T'} = \sqrt{\left\vert \frac{Q^2}{\qv^2}\right\vert +
\tan^2\frac{\theta}{2}} \tan\frac{\theta}{2}
\label{vTp}
\eeq
are the usual lepton factors,  $\theta$ is the electron scattering angle 
and $Q^2=\omega^2-\qv^2 < 0$ is the space--like four momentum transfer 
of the vector boson carrying the electromagnetic 
($\gamma$) or the weak neutral ($Z_0$) interaction.

In (\ref{asym}) the nuclear and nucleon's structure are embedded in both
the parity conserving (electromagnetic) longitudinal and transverse  
($R^L$ and $R^T$) response functions and in the parity violating
(weak neutral) ones. These classify as well in vector longitudinal and
transverse ($R^L_{AV}$, $R^T_{AV}$) and axial transverse $R^{T'}_{VA}$
responses, the first (second) index in the subscript referring to the
vector/axial nature of the weak neutral leptonic (hadronic) current.
Finally the scale of the asymmetry is set by
\beq
{\cal A}_0 = \frac{\sqrt{2}G m^2_\N}{\pi\alpha}
\frac{|Q^2|}{4m_\N^2} \approx 6.5\times10^{-4}\tau \qquad
\left(\tau=\frac{|Q^2|}{4m_\N^2} \right)
\label{A0}
\eeq
in terms of the electromagnetic ($\alpha$) and Fermi  ($G$) 
coupling constants ($m_\N$ is the nucleon mass).

To gain informations on the structure of nuclei and nucleons the 
standard Coulomb, electric and magnetic multipole decomposition
\beq
R^L(q,\omega) = \sum_{J\ge 0} F_{CJ}^2 (q)
\label{RL}
\eeq
and
\beq
R^T(q,\omega) = \sum_{J\ge 1}\left\{ F_{EJ}^2(q) + 
F_{MJ}^2(q) \right\}
\label{RT}
\eeq
for the parity conserving (electromagnetic) and
\bea
R^L_{AV}(q,\omega) &=& a_A\sum_{J\ge 0} F_{CJ}(q){\widetilde F}_{CJ}(q)
\label{RLAV}\\
R^T_{AV}(q,\omega) &=& a_A\sum_{J\ge 1} 
\left\{ F_{EJ}(q){\widetilde F}_{EJ}(q) + 
F_{MJ}(q){\widetilde F}_{MJ}(q)\right\}
\label{RTAV}\\
R^{T'}_{VA}(q,\omega) &=& -a_V\sum_{J\ge 1} 
\left\{ F_{EJ}(q){\widetilde F}_{MJ_5}(q) + 
F_{MJ}(q){\widetilde F}_{EJ_5}(q)\right\}
\label{RTPAV}
\eea
for the parity violating (weak neutral) responses are performed.

In the above formulas  $\omega$ is supposed to be given so the 
responses actually
become functions of $q$ only. Furthermore the Standard Model for the
electroweak interaction is assumed: thus the vector and axial--vector
leptonic coupling at the tree level read
\bea
a_V &=& 4\sin^2\theta_W - 1
\label{av}\\
a_A &=& -1
\label{aa}
\eea
in terms of the Weinberg's angle. Finally the form factors, chosen to
be real, with (without) a tilde are the weak--neutral (electromagnetic) ones.
They, of course, split into isoscalar and isovector components according
to the isospin decomposition
\beq
J_{\mu}= \xi^{(0)}\left(J_\mu\right)_{00} +
\xi^{(1)}\left(J_\mu\right)_{10} 
\label{current}
\eeq
of the hadronic current.

Again, on the basis of the Standard Model at tree level, one has
\beq
\xi^{(0)}=\xi^{(1)} = 1
\label{csiem}
\eeq
in the electromagnetic sector. In the weak neutral sector, instead, 
one has
\bea
\xi^{(0)}&=&\beta_V^{(0)}= -2\sin^2\theta_W \simeq -0.461,
\label{csiw1}\\
\xi^{(1)}&=&\beta_V^{(1)}= 1-2\sin^2\theta_W \simeq 0.538
\label{csiw2}
\eea
for the vector coupling and
\beq
\xi^{(0)}=\beta_A^{(0)}=0, \qquad 
\xi^{(1)}=\beta_A^{(1)}=1 
\label{csiwax}
\eeq
for the axial one.

All the above formulas hold valid in the approximation of exchanging only one
vector boson between the lepton and the hadron and up to a possible parity
admixture in the nucleon itself (anapole moment of the nucleon) or in
the nuclear states.The latter would  stem from parity--violating components
in the nucleon--nucleon interaction. These items will not be dealt with here.
For a comprehensive treatment we refer the interested reader to the specialized
literature on the subject.

\section{Elastic polarized electron scattering from spin zero, 
isospin zero nuclei}

As pointed out long time ago by Feinberg and Walecka\cite{Feen}
formula (\ref{asym}) applied 
to the elastic scattering of polarized electrons from a spin--zero, 
isospin--zero nuclear target leads to the simple expression
\beq
{\cal A} = {\cal A}_0  2 \sin^2\theta_W
\label{asymel}
\eeq
for the asymmetry.

It thus seemed that the opportunity was there to address the physics of the
Standard Model in the low energy regime, being  (\ref{asymel}) a 
``model--independent'' expression.

However  things are not so simple because the
isospin purity of the nuclear states, on which (\ref{asymel}) relies, 
is not realized in nature: indeed the proton and neutron quantum states 
are different in a nucleus.

Accordingly, to account for the isospin breaking, 
 (\ref{asymel}) should be recast as follows
\bea
{\cal A}&&={\cal A}_0 a_A\beta_V^{(0)}
\left\{ {\displaystyle
\frac {1+ {\displaystyle\frac{\beta_V^{(1)}}{\beta_V^{(0)}}} 
{\displaystyle\frac{<J_i |{\hat M}_{0;10}| J_i>}
{<J_i |{\hat M}_{0;00}| J_i> }} }
{1+ {\displaystyle\frac{<J_i |{\hat M}_{0;10}| J_i>}
{<J_i |{\hat M}_{0;00}| J_i> }} }}
\right\}
\label{nuclasym}\\
&&\simeq {\cal A}_0 2\sin^2\theta_W \left\{
1 + \left(\frac{\beta_V^{(1)}}{\beta_V^{(0)}} -1\right)
{\displaystyle \frac
{<J_i |{\hat M}_{0;10}| J_i> }{<J_i |{\hat M}_{0;00}| J_i>} } 
\right\}
\nonumber \\
&&= {\cal A}_0 2\sin^2\theta_W \left\{
1 - \frac{1}{2\sin^2\theta_W}
{\displaystyle \frac
{<J_i |{\hat M}_{0;10}| J_i> }{<J_i |{\hat M}_{0;00}| J_i>} } 
\right\}\, .
\nonumber
\eea

The above formula should of course be used ``cum grano salis''. 
Indeed the expansion in the arguably small ground state matrix element 
of the {\it isovector} monopole operator ${\hat M}_{0;10}$ is warranted
except where the ground state matrix element of the {\it isoscalar} monopole
operator ${\hat M}_{0;00}$ 
is also or vanishing or very small, which
happens, of course, at or close to the diffraction minima of the elastic
cross--sections.

Thus, barring for these small domains, one would like to estimate,
in the formula for the asymmetry
\beq
{\cal A}={\cal A}_0 2\sin^2\theta_W \left[ 1 + \Gamma(q)\right]\ ,
\label{asym2}
\eeq
the impact of the nuclear dependent term  $\Gamma(q)$,  whose definition 
follows by comparing  (\ref{nuclasym}) with  (\ref{asym2}).   

Before specifically addressing this issue, let us first answer the question:
given that $\Gamma(q)$ is not zero, is it still small enough to render worth 
trying a measurement of $\sin^2\theta_W$  with parity violating electron 
scattering?

Possibly this was the case a decade ago. Indeed around 1990 the value 
of the Weinberg's angle quoted in the literature read
\beq
\sin^2 \theta_W = 0.227\pm 0.005 \, ,
\label{Wein}
\eeq
namely it was given with a precision of 2.2\%.

Accordingly in such a condition a meaningful determination of 
$\sin^2\theta_W$  would have required on the one hand a measurement 
of the asymmetry to an accuracy of $1\div 2\%$ and on the other to 
assume the isospin impurity in a nucleus to be below such a level.

Let us now see how these figures translate into a kinematical constraint.
 Given that an
accuracy of $10^{-7}$ could be reached in measuring ${\cal A}$, a test 
of the Standard Model, as expressed by (\ref{Wein}),
would have had a good chance to be performed 
providing ${\cal A}\ge 10^{-5}$, in turn implying [from (\ref{A0})] 
$q \ge 1.75$~fm$^{-1}$. Now, always from (\ref{A0}), it follows
that ${\cal A}$ grows with $q$, but, at the same time, the elastic form
factor falls off with $q$: one would thus choose the range 
\beq
1.75  \le q  \le 3.5 \,{\mathrm fm}^{-1} .
\label{range}
\eeq
as a reasonnable compromise between these two opposite requirements.

The question to be addressed was (and is) then: how large is $\Gamma (q)$
in the range  (\ref{range})~? Is there  $\Gamma(q) \le 10^{-2}$~?
Whatever the answer to these questions might be, today the above arguing
is of course untenable since now
\beq
\sin^2\theta_W   =  0.23055  \pm 0.00041 \, ,
\label{Weinlast}
\eeq
i.e, the Weinberg's angle is known with an accuracy of $0.18\%$.

Yet for others relevant observables, for example the strange parity content
of the nucleon, although a generalization of the expression  (20) for the
asymmetry is required, a reliable theoretical handling of $\Gamma (q)$ is still 
crucial. This issue will be addressed in the next Section.

\section{A simple model for isospin breaking}

The isospin symmetry is broken in atomic nuclei by the Coulomb force which
pushes the protons orbits outward with respect to the neutrons ones.
On the other hand the strong proton--neutron
interaction tries to equalize the proton's and neutron's Fermi energies, thus
acting in the direction of restoring the isospin symmetry: the balance between
these two effects is believed to leave a generally modest symmetry breaking
in nuclei,which thus lends itself to a perturbative treatment.

To explore this physics Donnelly {\it et al.}\cite{Donn} worked out a simple
model characterized by two monopole states ($J^\pi =0^+$), one isoscalar and 
one isovector, mixed by the isospin
breaking interaction, in particular the Coulomb one. In this scheme the
dominantly  isospin $T = T_0$ ground state is actually represented by the
superposition 
\beq
|``T_0''> = \cos\chi\,|T_0> + \sin\chi\, |T_0+1>\, ,
\label{T0}
\eeq
and the dominantly  $T_1 = T_0 + 1$ excited state by the orthogonal 
combination
\beq
|``T_1''> = -\sin\chi\,|T_0> + \cos\chi\, |T_0+1>\, .
\label{T1}
\eeq

The associated ground state matrix elements of the isoscalar and isovector
monopole Coulomb operators read then, respectively,
\bea
&&<``T_0''|{\hat M}_{0;00}(q)|``T_0''>
\nonumber \\
&&\qquad = \cos^2\chi\,<T_0|{\hat M}_{0;00}|T_0> +
\sin^2\chi\, <T_0+1|{\hat M}_{0;00}|T_0+1>
\label{M00me}
\eea
and
\bea
&&<``T_0''|{\hat M}_{0;10}(q)|``T_0''> = 
\cos^2\chi\,<T_0|{\hat M}_{0;10}|T_0> 
\label{M10me} \\
&& + \sin^2\chi\, <T_0+1|{\hat M}_{0;10}|T_0+1> +
2\sin\chi\cos\chi\, <T_0+1|{\hat M}_{0;10}|T_0>\, .
\nonumber
\eea

Let us then  consider $N = Z$ nuclei. The Clebsch--Gordon (CG) 
coefficients entering into the reduction of the above matrix elements 
in isospace are 1 and $1/\sqrt{3}$ , respectively, in formula (\ref{M00me}), 
whereas in  (\ref{M10me}) only the third term survives and the associated 
 CG is $1/\sqrt{3}$.

One thus obtains for the asymmetry in leading order of the mixing
angle $\chi$ the expression
\bea
{\cal A} &&= {\cal A}_0 a_A {\displaystyle\frac
{\beta_V^{(0)}<0^+;0\|{\hat M}_{0;00}\|0^+;0> +
\beta_V^{(1)}2\chi\frac{1}{\sqrt{3}}<0^+;1\|{\hat M}_{0;10}\|0^+;0>}
{ <0^+;0\|{\hat M}_{0;00}\|0^+;0> +
2\chi\frac{1}{\sqrt{3}}<0^+;1\|{\hat M}_{0;10}\|0^+;0>} }
\nonumber \\
&&\quad \simeq {\cal A}_0 2\sin^2\theta_W \{1 +\Gamma(q)\}
\label{asym3}
\eea
where
\beq
\Gamma(q) = 2\left(\frac{\beta_V^{(1)}}{\beta_V^{(0)}} - 1\right)
\chi {\cal R}(q) = 
-\frac{1}{\sin^2\theta_W}\chi {\cal R}(q)
\label{Gamma}
\eeq
and
\beq
{\cal R}(q) = \frac{1}{\sqrt{3}}
\frac{<0^+;1\|{\hat M}_{0;10}\|0^+;0>}{<0^+;0\|{\hat M}_{0;00}\|0^+;0>}
= \frac{F_{C0}(0^+;``1'';``0''\to 0^+)}
{F_{C0}(0^+;``0'';``0''\to 0^+)}\, .
\label{Erre}
\eeq
In the above the double bar matrix elements are meant to be reduced
in isospace.
Note that ${\cal R}(q)$ simply represents the ratio between the inelastic 
and the elastic form factors associated with the two monopoles states of 
our $N = Z$ nucleus.

Explicit calculations of $\Gamma(q)$ for a few nuclei have been performed
by Donnelly {\it et al.}\cite{Donn} in a Wood--Saxon single particle wave
functions basis with various effective interactions and in different
configurations spaces. 
We display in Fig.~1a and 1b their results for $\Gamma(q)$ in C$^{12}$ 
and Si$^{28}$. The singularities in the curves should
be disregarded since they don't have, as previously discussed, any physical
significance. Although the results of ref.\cite{Donn} are model dependent 
(they do change significantly according to the effective 
interaction employed), 
yet they convey the message that for ligth $N = Z$ nuclei like C$^{12}$ 
the isospin breaking remains tiny indeed, below 1\% over the whole range 
(\ref{range}) of $q$.

\begin{figure}[t]
\mbox{\epsfig{file=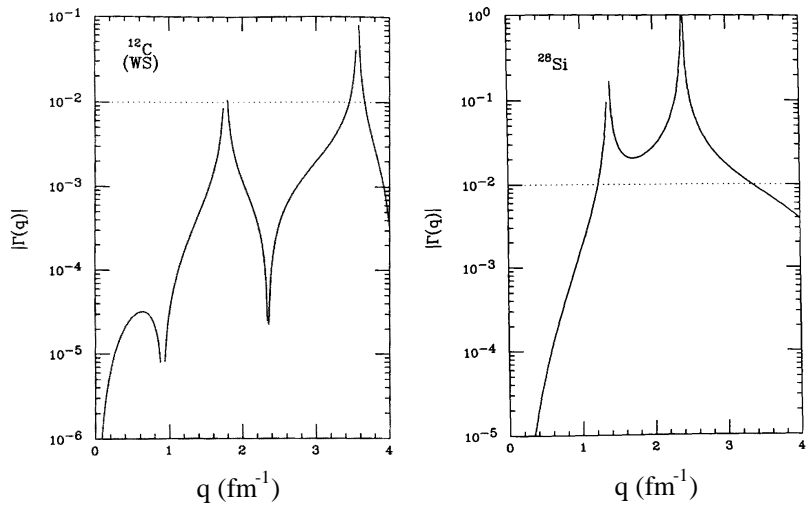,width=0.9\textwidth}}
\vskip 0.1cm
\caption[Fig.~\ref{fig1}]{\label{fig1}
The nuclear--structure--dependent part of the parity--violating 
asymmetry as defined by eqs,~(\ref{Gamma}) and (\ref{Erre}). 
{\it Left:} calculations
for elastic scattering from $^{12}$C using Woods--Saxon single--particle 
wave functions. The dotted line at $|\Gamma(q)|=10^{-2}$ indicates the 
level above which the structure--dependent effects would have confused the
interpretation of the asymmetry as a test of electroweak theories, when 
these were known with the precision given by (\ref{Wein}).
{\it Right:} the same as in the left panel, but for $^{28}$Si. For this 
nucleus Harmonic Oscillator single--particle wave functions and the 
shell model amplitudes as given by W.C. Haxton (unpublished) have been used.
}
\end{figure}         

Note also that in reaching this result  the
mixing angle $\chi$ has been extracted,
with a quite conservative attitude,  from the perturbative formula
\beq
\sin^2\chi = \frac{\left|<T_0+1|H_{CSV}|T_0>\right|^2}
{\left(E_{T_0+1}-E_{T_0}\right)^2}
\label{sinchi}
\eeq
( $H_{CSV}$ is the charge symmetry violating part of the nuclear 
hamiltonian). In fact, while the experiments 
indicate for the matrix element appearing in the numerator of (\ref{sinchi}) 
a value ranging, in C$^{12}$, between 150 and 300~keV, the latter value
has been adopted in  obtaining Fig.~1a and 1b (moreover 
$\Delta E = E_{T_0+1}-E_{T_0}\simeq 17$~MeV according to the simple two
levels model). 

 These findings are supported by the results of a quite sophisticated
calculation of Ramavataram {\it et al.}\cite{Ramava}. These authors, 
using a state of the art variational wave function originally due to 
Pandharipande {\it et al.}\cite{Pandha}, obtain in He$^4$ a 
$\Gamma(q)$ always well below 1\% in the whole range (\ref{range}). 
Of course this outcome also reflects the particularly rigid structure 
of He$^4$, whose first excited state lies at 20.1~MeV (the first excited 
state of C$^{12}$ is at 4.44~MeV).

It should however be pointed out that, for $N = Z$  but
heavier  nuclei like Si$^{28}$, the isospin breaking, while small, grows and
reaches a few \% in the range of q given by (\ref{range}).

\section{The case of  $N \ne Z$ nuclei}

For nuclei with $N \ne Z$ a  novel (and important) 
feature appears. Although it will be illustrated in the specific case 
of nuclei with $N = Z + 2$, and therefore with third isospin 
component $M_T = -1$, it remains valid in all cases.

 Sticking always to the two levels model\cite{Donn} 
we consider then a nucleus with a dominant  isospin $T_0 = 1$ component
in the ground state. Let in addition the nucleus have an excited state with
a dominant isospin $T_1 = T_0 + 1 = 2$. Both states are further 
characterized by having $J^\pi  = 0^+$.

Proceeding as in the case of a  $Z = N$ nucleus, one arrives to an asymmetry
still given by an expression like (\ref{asym3}), but with the model 
dependent nuclear term  $\Gamma(q)$ reading now as follows 
\bea
&&\Gamma(q) = \frac{1}{2}\beta_V^{(1)}
\left\{
{-\frac{1}{\sqrt{6}}<0^+;1\|{\hat M}_{0;1}\|0^+;1> +
2\sqrt{\frac{1}{10}}\chi <0^+;2\|{\hat M}_{0;1}\|0^+;1> }
\right\}
\nonumber\\
&&\qquad\quad\times\left\{
\sqrt{\frac{1}{3}} <0^+;1\|{\hat M}_{0;0}\|0^+;1>  
-\frac{1}{\sqrt{6}} <0^+;1\|{\hat M}_{0;1}\|0^+;1> +
\right.
\nonumber\\
&&\qquad\qquad\qquad\qquad \left.
+ 2\sqrt{\frac{1}{10}}\chi <0^+;2\|{\hat M}_{0;1}\|0^+;1> 
\right\}^{-1}
\label{Gamma1}\\
\eea
which can again be recast according to
\beq
\Gamma(q) = \frac{1}{2}\beta_V^{(1)}
{\displaystyle\frac 
{F_{C0}(0^+;``1'';``1''\to 0^+)_{\mathrm isovector}}
{F_{C0}(0^+;``1'';``1''\to 0^+)_{\mathrm total}} }\, .
\label{Gamma2}
\eeq
Namely $\Gamma(q)$ turns out to be, like before, proportional to the 
ratio between  the inelastic (isovector) and the elastic (this time both 
isoscalar and isovector) form factors.

Now from (\ref{Gamma1}) a new feature is immediately apparent: unlike the 
$N = Z$ case, where $\Gamma(q)$ was found to be proportional to 
the mixing parameter $\chi$, here $\Gamma(q)$, in addition to terms 
proportional to $\chi$, also embodies terms {\it independent} from it. 
As a consequence, $\Gamma(q)$ turns out to be now much larger,
 as it is clearly observed  in Fig.~2a and 2b, 
where the results of ref.\cite{Donn} are displayed for C$^{14}$ and 
Si$^{30}$.

\begin{figure}[t]
\mbox{\epsfig{file=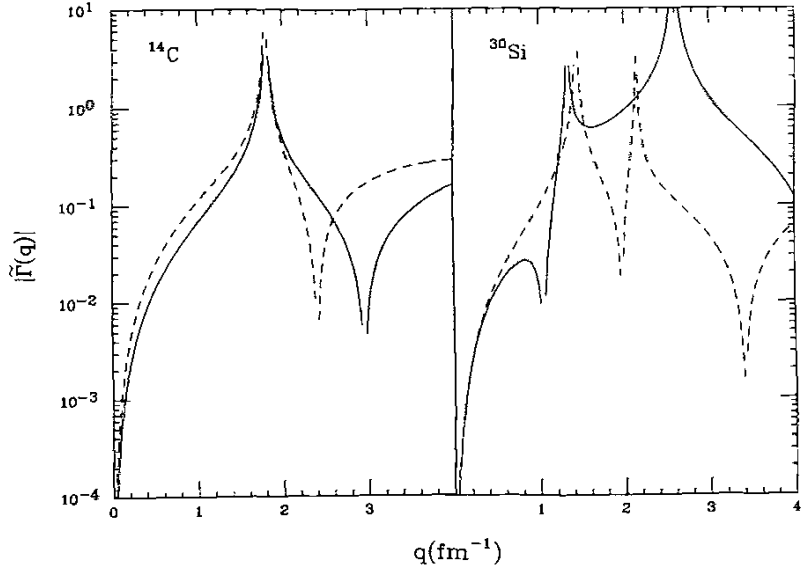,width=0.9\textwidth}}
\vskip 0.1cm
\caption[Fig.~\ref{fig2}]{\label{fig2}
The structure--dependent part of the parity--violating asymmetry for
elastic scattering as defined in eqs.~(\ref{tildegam}) and (\ref{betap}).
The results for $^{14}$C using a 1p--shell model space (solid line) and a
$2\hbar\omega$--space (dashed line) are displayed. Also displayed are the
results for $^{30}$Si in the extreme single--particle shell model (solid 
line) and in a full 2s1d shell--model calculation (dashed line). No isospin
mixing effects are included.
}
\end{figure}         

Actually what is shown in the figures 2a and 2b is not $\Gamma(q)$, but 
 ${\widetilde\Gamma} (q)$ since   the  $N \ne Z$  nuclei 
are more appropriately discussed in terms of protons and neutrons rather 
than in term of isospin and, accordingly, the expression (\ref{asym2}) 
for the asymmetry is now more conveniently
recast into the form:
\beq
{\cal A}={\cal A}_0 a_A\left(\beta_V^p+\frac{N}{Z}\beta_V^n\right)
\left[ 1 + {\widetilde\Gamma}(q)\right]
\label{asym4}
\eeq
where
\bea
\beta_V^p &=& \frac{1}{2}\left(\beta_V^{(0)} +\beta_V^{(1)}\right)
=0.038\, ,
\label{betavp}\\
\beta_V^n &=& \frac{1}{2}\left(\beta_V^{(0)} -\beta_V^{(1)}\right)
= -\frac{1}{2}
\label{betavn}
\eea
and
\beq
{\widetilde\Gamma}(q) = \frac{1}{2}{\tilde\beta}'_V
{\displaystyle\frac 
{\left\langle 0^+\left|\frac{1}{2}\left(1-\frac{Z}{N}\right){\hat M}_{0;00}
+\frac{1}{2}\left(1+\frac{Z}{N}\right){\hat M}_{0;10}\right|0^+ \right\rangle}
{<0^+\left| {\hat M}_{0;00} + {\hat M}_{1;10}\right|0^+>} }\, ,
\label{tildegam}
\eeq
being
\beq
{\tilde\beta}'_V = 4\frac{N}{Z}\frac{\beta_V^n}{\beta_V^p + 
{\displaystyle\frac{N}{Z}}\beta_V^n}\, .
\label{betap}
\eeq
It is immediately checked that by setting $N = Z$ in the above formulas 
one gets ${\widetilde\Gamma}(q)\to \Gamma(q)$ and ${\tilde\beta}'_V =
1/\sin^2\theta_W$.

Let us further observe that, although in the figures 2a and 2b 
${\widetilde\Gamma}(q)$ 
appears indeed to be markedly model dependent, yet the basic message 
previously anticipated clearly stands out:
in the range of momentum transfers (\ref{range}) ${\widetilde\Gamma}(q)$ 
assumes values typically ranging  between  10 and 50\%.

In conclusion, from the previous two Sections it follows that
 for light $N = Z$ nuclei, like He$^4$ and C$^{12}$,
\begin{itemize}
\item{} 
a test of the electroweak theory with parity--violating polarized elastic
electron scattering experiments is out of question or, perhaps, 
only marginally, if at all, possible at very low momentum transfer 
in He$^4$;

\item{} 
however  many investigations show that 
elastic (and, also, quasielastic, not addressed here)
parity violating polarized electron scattering experiments 
can be usefully exploited to 
unravel the strange  form factor of the nucleon.
\end{itemize}

For medium--heavy and heavy nuclei $\Gamma(q)$ grows with the mass number 
$A$, especially when $N \ne Z$ (large terms not proportional to 
$\chi$ appear!). 
Therefore parity--violating polarized electron scattering can be 
advantageously used to measure the amount of isospin breaking in a nucleus 
or, better yet, the neutron distribution.

\section{The neutron distribution}

To appreciate how the neutron distribution can be measured with polarized
elastic electron scattering it helps to revisit the previous concepts with a
somewhat different language. Let us thus observe that for $T = 0$ nuclei 
($N = Z$) if isospin is a ``good symmetry'' then the standard unpolarized 
electron scattering measures the {\it isoscalar} nuclear density. 
In these conditions the latter also fixes  the polarized elastic 
electron scattering which is thus independent from any further nuclear 
structure information.

If, however, isospin is sligthly broken, then the {\it isovector} nuclear 
density also enters, as a small perturbation, in the elastic 
polarized electron 
scattering thus introducing an additional model dependence.

Quite on the contrary, for $T \ne 0$  ($N \ne Z$) nuclei 
 both the {\it isoscalar} and the {\it isovector} 
densities enter into the scattering process, no matter if isospin is or 
is not a perfect symmetry. In this instance, as already pointed out, it 
is preaferable to use the neutron--proton language. Accordingly the 
ground state matrix element of the Coulomb monopole operator
\beq
<0^+\left|{\hat M}_0(q)\right|0^+> =
\frac{1}{\sqrt{4\pi}}\int d\xv j_0(qx)\rho(\xv)\, ,
\label{monopole}
\eeq
where  $j_0$ is the zeroth order spherical Bessel function and  
$\rho(\xv)$ is the matter density of the nucleus, will display both an  
isoscalar and an isovector components, according to the expressions
\beq
<0^+\left|{\hat M}_{0;00}(q)\right|0^+> =
\frac{1}{\sqrt{4\pi}}\int d\xv j_0(qx)
\frac{\rho_p(\xv)+\rho_n(\xv)}{2}
\label{isosca}
\eeq
and
\beq
<0^+\left|{\hat M}_{1;00}(q)\right|0^+> =
\frac{1}{\sqrt{4\pi}}\int d\xv j_0(qx)
\frac{\rho_p(\xv)- \rho_n(\xv)}{2}
\label{isovec}
\eeq
In the above $\rho_p$ and $\rho_n$ are the protons and neutrons densities, 
respectively.

When (\ref{isosca})  and (\ref{isovec}) are inserted into the formula 
(\ref{asym4}) it is then an easy matter to obtain for the asymmetry the 
expression\footnote{Note that,  by comparison, (\ref{Gamma3}) yields
\beq
{\widetilde\Gamma}(q) = \frac{N}{Z} \frac{\beta_V^n}{\beta_V^p +
{\displaystyle\frac{N}{Z}}\beta_V^n}
\left[ 1 -
{\displaystyle\frac
{Z \int d\xv j_0(qx)\rho_n(\xv)}
{N \int d\xv j_0(qx)\rho_p(\xv)} } \right]
\, .
\label{Gamma3}
\eeq
}
\beq
{\cal A} = {\cal A}_0 a_A \left\{ \beta_V^p +
\beta_V^n {\displaystyle\frac
{\int d\xv j_0(qx)\rho_n(\xv)}
{\int d\xv j_0(qx)\rho_p(\xv)} } \right\}
\label{asym5}
\eeq

Since, according to the Standard model, 
\beq
\beta_V^p = 0.038  \qquad{\mathrm and}\quad  \beta_V^n = - 0.5 \,,
\label{boh}
\eeq
(\ref{Gamma3}) shows that {\it the asymmetry in the parity--violating 
elastic polarized electron scattering represents an almost direct measurement 
of the Fourier transform of the neutron density}, the analogous 
quantity for the protons being fixed by the elastic unpolarized electron
scattering.

In particular a rigorous $q$--independence of ${\cal A}/{\cal A}_0$ 
would imply
\beq
Z \rho_n(\xv) = N \rho_p(\xv),
\eeq
i.e. pure isospin symmetry! Of course, as previously discussed, in nuclei
the distribution of neutrons differs from the one of protons.

To gain a first insight on how this difference would be perceived 
in a parity violating elastic electron scattering experiment
Donnelly {\it et al.} have computed the asymmetry (\ref{asym5}) and the
${\widetilde\Gamma}(q)$ of eq.~(\ref{Gamma3}) for Ca$^{40}$, Ca$^{48}$ and 
Pb$^{208}$ using
phenomenological proton densities which well accomplish for the elastic
unpolarized electron scattering.
For example, for Ca$^{40}$ they employ the
well--known 3 parameters Fermi distribution:
\beq
\rho(\rv)=\rho_0{\displaystyle\frac{1+\omega{\displaystyle\frac{r^2}{R^2}}}
{1+e^{{\displaystyle (r-R)/a}}} }\, .
\label{Fermi3}
\eeq
For the neutrons they use the same densities, 
euristically enlarging however, for explorative purposes, 
the radius parameter by 0.2~fm. 
Their results are displayed in Fig.~3. To grasp the significance of 
this figure it helps to expand ${\widetilde\Gamma}(q)$ as follows:
\beq
{\widetilde\Gamma}(q) = \frac{\beta_V^n}{\beta_V^p +
{\displaystyle\frac{N}{Z}}\beta_V^n}
\left[ \frac{N}{Z} - {\displaystyle\frac
{N - \frac{1}{6}q^2 R^2_n} 
{ \int d\xv j_0(qx)\rho_p(\xv)} } + \dots \right]
\, .
\label{Gamma4}
\eeq
It is thus clear that the region around the first dip/peak carries 
information on the neutron radius whereas the fourth moment of the neutron
distribution will be observed nearby the second dip/peak and so on.
On the basis of this expansion one finds that  in the case
of Pb$^{208}$ at $q\simeq 0.5$~fm$^{-1}$,  with 
${\cal A}\simeq 8\times 10^{-8}$, a 1\%  change in the neutron radius 
is reflected in a change of about 6\% in the asymmetry.

\begin{figure}[t]
\mbox{\epsfig{file=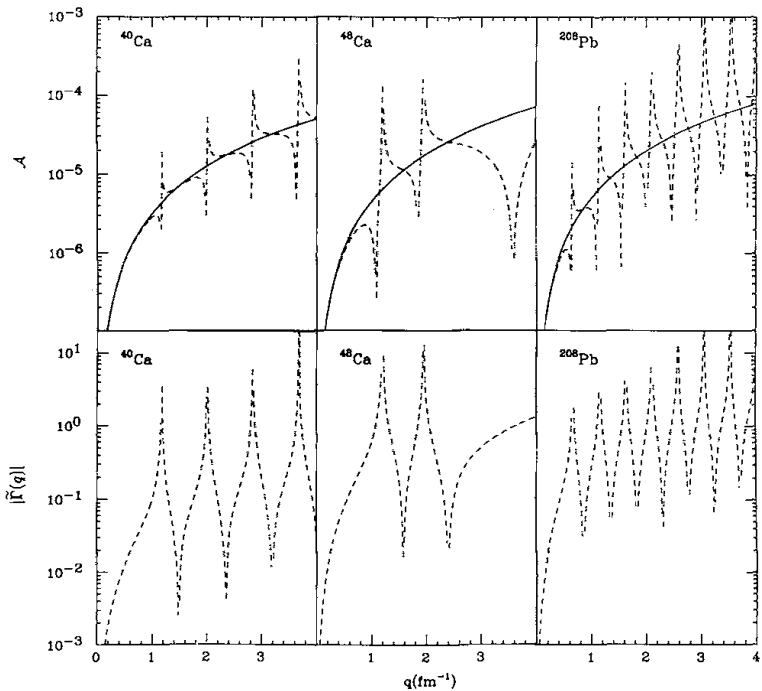,width=0.9\textwidth}}
\vskip 0.1cm
\caption[Fig.~\ref{fig3}]{\label{fig3}
The parity--violating asymmetry (upper row) and the structure--dependent 
part of the asymmetry as defined in eq.~(\ref{Gamma4}) (lower row) for
elastic electron scattering from $^{40}$Ca, $^{48}$Ca and $^{208}$Pb.
Displayed are the calculations with $\rho_n(r)/N=\rho_p(r)/Z$ [solid line,
corresponding to ${\widetilde\Gamma}(q)=0$] and for 
$\rho_n(r)/N\ne \rho_p(r)/Z$ (dashed line). The density parameterizations
are discussed in the text and specified in ref.\cite{Donn}.
}
\end{figure}         

It might be interesting to observe that formula (\ref{Gamma4}) can be 
generalized in the sense that the Fourier transform of the pure Fermi
distribution can be, although not exactly, quite accurately analytically
expressed. Indeed for this density the form factor turns out to 
be\cite{noi1}
\beq
F_{\mathrm Fermi}(q) =\rho_0 (2\pi)^2R^2 a
\left\{j_1(qR){\tilde y}_0(\pi qa)
- {\tilde j}_1(\pi qa)y_0(qR)\right\}
\frac{\pi qa}{{\tilde j}_0(\pi qa)}
\label{FForm}
\eeq
where the $j(y)$ and the ${\tilde j}({\tilde y})$ are the ``spherical''
and the ``modified spherical'' Bessel functions of first (second) kind,
respectively \cite{Abram}.

The form factor $F_3$ of the three--parameters Fermi distribution is
then expressed in terms of $F_{\mathrm Fermi}$ according to:
\beq
F_3(q) \simeq F_{\mathrm Fermi}(q) -\frac{\omega}{R^2}
\frac{d^2}{dq^2}\left[ q F_{\mathrm Fermi}(q)\right]\, .
\label{F3}
\eeq
When inserted into (\ref{asym5}) and (\ref{Gamma3}), the above formula
allows to analytically recover the results of Donnelly for Ca$^{40}$.

\section{Flaws and hopes}

The results of the previous section are flawed essentially 
by two shortcomings:
one relates to the factorization of the single nucleon physics, which
has been assumed in deducing (\ref{asym5}) from (\ref{asym4}).
Especially when $q$ is large, and thus relativistic effects become
substantial, such a procedure almost certainly becomes unwarranted.
It is clear that this point must be more carefully addressed in future
research.

The second problem relates to the approximation of considering just
the Fourier transform of the charge and neutron distributions,
which  corresponds to consider a single vector boson exchange 
between the impinging field and the target, leaving the electron to be
described by plane waves. This is clearly insufficient, especially in 
heavy nuclei, where, in fact, the electron wave is quite distorted
by the nuclear Coulomb field. To account for this effect a heavy 
computational effort is presently carried out by an MIT--Indiana 
University collaboration. A more modest, but perhaps also useful 
approach, resorts to a kind of eikonal approximation to describe the 
distortion of the electron wave.

We feel confident that these difficulties will in the end be overcome.
It will thus become possible to measure the neutron distribution 
in the ground state of atomic nuclei with parity violating {\it elastic}
scattering experiments. 

This most remarkable occurrence goes in parallel with the finding
of Alberico {\it et al.}\cite{noi2} for the {\it quasi--elastic} polarized 
electron--nucleus scattering. Indeed these authors show that, being
the $Z_0$ almost ``blind'' to protons, the PV longitudinal response
of an uncorrelated system of protons and neutrons, like the relativistic
Fermi gas, to a polarized beam of electrons is almost vanishing. 
Departure from this expectation will thus signal the effect of 
neutron--proton correlations in nuclei, which are expected to be 
especially relevant in the isoscalar channel.

Thus polarized electron scattering experiments appear  to open a 
window on crucial, and till now insufficiently explored, aspects of the 
nuclear structure.

A final remark is in order:  parity--violating 
{\it nuclear} electron scattering experiments, initially conceived as 
a tool for exploring the Standard Model at the nuclear level, will 
turn out, in the end, to represent a tool for providing the information
the {\it atomic} parity violating experiments need to accurately test the
Standard Model at the atomic level: namely the neutron distribution. Indeed 
the precision on the measured energy levels of the Cesium
atoms, which is required to test the Standard Model, is so high that 
it cannot be reached without controlling also the neutron distribution 
in nuclei.


\end{document}